\documentclass{article}
\usepackage{amssymb}
\makeatletter
\newcount\@hour\newcount\@minute\chardef\@x10\chardef\@xv60
\def\tcitime{
\def\@time{%
  \@minute\time\@hour\@minute\divide\@hour\@xv
  \ifnum\@hour<\@x 0\fi\the\@hour:%
  \multiply\@hour\@xv\advance\@minute-\@hour
  \ifnum\@minute<\@x 0\fi\the\@minute
  }}%

\@ifundefined{hyperref}{}{}

\@ifundefined{qExtProgCall}{\def\qExtProgCall#1#2#3#4#5#6{\relax}}{}
\def\QCTOpt[#1]#2{%
  \def\QCTOptB{#1}
  \def\QCTOptA{#2}
}
\def\QCTNOpt#1{%
  \def\QCTOptA{#1}
  \let\QCTOptB\empty
}
\def\Qct{%
  \@ifnextchar[{%
    \QCTOpt}{\QCTNOpt}
}
\def\QCBOpt[#1]#2{%
  \def\QCBOptB{#1}
  \def\QCBOptA{#2}
}
\def\QCBNOpt#1{%
  \def\QCBOptA{#1}
  \let\QCBOptB\empty
}
\def\Qcb{%
  \@ifnextchar[{%
    \QCBOpt}{\QCBNOpt}
}
\def\PrepCapArgs{%
  \ifx\QCBOptA\empty
    \ifx\QCTOptA\empty
      {}%
    \else
      \ifx\QCTOptB\empty
        {\QCTOptA}%
      \else
        [\QCTOptB]{\QCTOptA}%
      \fi
    \fi
  \else
    \ifx\QCBOptA\empty
      {}%
    \else
      \ifx\QCBOptB\empty
        {\QCBOptA}%
      \else
        [\QCBOptB]{\QCBOptA}%
      \fi
    \fi
  \fi
}
\newcount\GRAPHICSTYPE
\GRAPHICSTYPE=\z@
\def\GRAPHICSPS#1{%
 \ifcase\GRAPHICSTYPE
   \special{ps: #1}%
 \or
   \special{language "PS", include "#1"}%
 \fi
}%
\def\graffile#1#2#3#4{%
    \leavevmode
    \raise -#4 \BOXTHEFRAME{%
        \hbox to #2{\raise #3\hbox to #2{\null #1\hfil}}}%
}%
\def\draftbox#1#2#3#4{%
 \leavevmode\raise -#4 \hbox{%
  \frame{\rlap{\protect\tiny #1}\hbox to #2%
   {\vrule height#3 width\z@ depth\z@\hfil}%
  }%
 }%
}%
\newcount\draft
\draft=\z@

\newif\ifwasdraft
\wasdraftfalse

\def\GRAPHIC#1#2#3#4#5{%
 \ifnum\draft=\@ne\draftbox{#2}{#3}{#4}{#5}%
  \else\graffile{#1}{#3}{#4}{#5}%
  \fi
 }%
\def\addtoLaTeXparams#1{%
    \edef\LaTeXparams{\LaTeXparams #1}}%
\newif\ifBoxFrame \BoxFramefalse
\newif\ifOverFrame \OverFramefalse
\newif\ifUnderFrame \UnderFramefalse

\def\BOXTHEFRAME#1{%
   \hbox{%
      \ifBoxFrame
         \frame{#1}%
      \else
         {#1}%
      \fi
   }%
}

\def\doFRAMEparams#1{\BoxFramefalse\OverFramefalse\UnderFramefalse\readFRAMEparams#1\end}%
\def\readFRAMEparams#1{%
 \ifx#1\end%
  \let\next=\relax
  \else
  \ifx#1i\dispkind=\z@\fi
  \ifx#1d\dispkind=\@ne\fi
  \ifx#1f\dispkind=\tw@\fi
  \ifx#1t\addtoLaTeXparams{t}\fi
  \ifx#1b\addtoLaTeXparams{b}\fi
  \ifx#1p\addtoLaTeXparams{p}\fi
  \ifx#1h\addtoLaTeXparams{h}\fi
  \ifx#1X\BoxFrametrue\fi
  \ifx#1O\OverFrametrue\fi
  \ifx#1U\UnderFrametrue\fi
  \ifx#1w
    \ifnum\draft=1\wasdrafttrue\else\wasdraftfalse\fi
    \draft=\@ne
  \fi
  \let\next=\readFRAMEparams
  \fi
 \next
 }%
\def\IFRAME#1#2#3#4#5#6{%
      \bgroup
      \let\QCTOptA\empty
      \let\QCTOptB\empty
      \let\QCBOptA\empty
      \let\QCBOptB\empty
      #6%
      \parindent=0pt%
      \leftskip=0pt
      \rightskip=0pt
      \setbox0 = \hbox{\QCBOptA}%
      \@tempdima = #1\relax
      \ifOverFrame
          \typeout{This is not implemented yet}%
          \show\HELP
      \else
         \ifdim\wd0>\@tempdima
            \advance\@tempdima by \@tempdima
            \ifdim\wd0 >\@tempdima
               \textwidth=\@tempdima
               \setbox1 =\vbox{%
                  \noindent\hbox to \@tempdima{\hfill\GRAPHIC{#5}{#4}{#1}{#2}{#3}\hfill}\\%
                  \noindent\hbox to \@tempdima{\parbox[b]{\@tempdima}{\QCBOptA}}%
               }%
               \wd1=\@tempdima
            \else
               \textwidth=\wd0
               \setbox1 =\vbox{%
                 \noindent\hbox to \wd0{\hfill\GRAPHIC{#5}{#4}{#1}{#2}{#3}\hfill}\\%
                 \noindent\hbox{\QCBOptA}%
               }%
               \wd1=\wd0
            \fi
         \else
            \ifdim\wd0>0pt
              \hsize=\@tempdima
              \setbox1 =\vbox{%
                \unskip\GRAPHIC{#5}{#4}{#1}{#2}{0pt}%
                \break
                \unskip\hbox to \@tempdima{\hfill \QCBOptA\hfill}%
              }%
              \wd1=\@tempdima
           \else
              \hsize=\@tempdima
              \setbox1 =\vbox{%
                \unskip\GRAPHIC{#5}{#4}{#1}{#2}{0pt}%
              }%
              \wd1=\@tempdima
           \fi
         \fi
         \@tempdimb=\ht1
         \advance\@tempdimb by \dp1
         \advance\@tempdimb by -#2%
         \advance\@tempdimb by #3%
         \leavevmode
         \raise -\@tempdimb \hbox{\box1}%
      \fi
      \egroup%
}%
\def\DFRAME#1#2#3#4#5{%
 \begin{center}
     \let\QCTOptA\empty
     \let\QCTOptB\empty
     \let\QCBOptA\empty
     \let\QCBOptB\empty
     \ifOverFrame 
        #5\QCTOptA\par
     \fi
     \GRAPHIC{#4}{#3}{#1}{#2}{\z@}
     \ifUnderFrame 
        \nobreak\par #5\QCBOptA
     \fi
 \end{center}%
 }%
\def\FFRAME#1#2#3#4#5#6#7{%
 \begin{figure}[#1]%
  \let\QCTOptA\empty
  \let\QCTOptB\empty
  \let\QCBOptA\empty
  \let\QCBOptB\empty
  \ifOverFrame
    #4
    \ifx\QCTOptA\empty
    \else
      \ifx\QCTOptB\empty
        \caption{\QCTOptA}%
      \else
        \caption[\QCTOptB]{\QCTOptA}%
      \fi
    \fi
    \ifUnderFrame\else
      \label{#5}%
    \fi
  \else
    \UnderFrametrue%
  \fi
  \begin{center}\GRAPHIC{#7}{#6}{#2}{#3}{\z@}\end{center}%
  \ifUnderFrame
    #4
    \ifx\QCBOptA\empty
      \caption{}%
    \else
      \ifx\QCBOptB\empty
        \caption{\QCBOptA}%
      \else
        \caption[\QCBOptB]{\QCBOptA}%
      \fi
    \fi
    \label{#5}%
  \fi
  \end{figure}%
 }%
\newcount\dispkind%

\def\makeactives{
  \catcode`\"=\active
  \catcode`\;=\active
  \catcode`\:=\active
  \catcode`\'=\active
  \catcode`\~=\active
}
\bgroup
   \makeactives
   \gdef\activesoff{%
      \def"{\string"}
      \def;{\string;}
      \def:{\string:}
      \def'{\string'}
      \def~{\string~}
    }
\egroup

\def\FRAME#1#2#3#4#5#6#7#8{%
 \bgroup
 \@ifundefined{bbl@deactivate}{}{\activesoff}
 \ifnum\draft=\@ne
   \wasdrafttrue
 \else
   \wasdraftfalse%
 \fi
 \def\LaTeXparams{}%
 \dispkind=\z@
 \def\LaTeXparams{}%
 \doFRAMEparams{#1}%
 \ifnum\dispkind=\z@\IFRAME{#2}{#3}{#4}{#7}{#8}{#5}\else
  \ifnum\dispkind=\@ne\DFRAME{#2}{#3}{#7}{#8}{#5}\else
   \ifnum\dispkind=\tw@
    \edef\@tempa{\noexpand\FFRAME{\LaTeXparams}}%
    \@tempa{#2}{#3}{#5}{#6}{#7}{#8}%
    \fi
   \fi
  \fi
  \ifwasdraft\draft=1\else\draft=0\fi{}%
  \egroup
 }%
\def\TEXUX#1{"texux"}

\long\def\QQQ#1#2{%
     \long\expandafter\def\csname#1\endcsname{#2}}%
\@ifundefined{QTP}{\def\QTP#1{}}{}
\@ifundefined{QEXCLUDE}{\def\QEXCLUDE#1{}}{}
\@ifundefined{Qlb}{}{}
\@ifundefined{Qlt}{}{}
\long\def\QQA#1#2{}%
\def\QTR#1#2{{\csname#1\endcsname #2}}
\def\EXPAND#1[#2]#3{}%
\def\NOEXPAND#1[#2]#3{}%
\def\LaTeXparent#1{}%
\def\ChildStyles#1{}%
\def\ChildDefaults#1{}%
\def\QTagDef#1#2#3{}%
\@ifundefined{StyleEditBeginDoc}{}{}
\def\QQfnmark#1{\footnotemark}

\@ifundefined{INDEX}{\def\INDEX#1#2{}{}}{}%
\@ifundefined{SUBINDEX}{\def\SUBINDEX#1#2#3{}{}{}}{}%
\@ifundefined{initial}%
   {\def\initial#1{\bigbreak{\raggedright\large\bf #1}\kern 2\p@\penalty3000}}%
   {}%
\@ifundefined{entry}{}{}%
\@ifundefined{primary}{}{}%
\@ifundefined{secondary}{}{}%
\@ifundefined{ZZZ}{}{\makeatletter\input gnuindex.sty\makeatother\makeindex\makeatletter}%
\@ifundefined{abstract}{%
 \def\abstract{%
  \if@twocolumn
   \section*{Abstract (Not appropriate in this style!)}%
   \else \small 
   \begin{center}{\bf Abstract\vspace{-.5em}\vspace{\z@}}\end{center}%
   \quotation 
   \fi
  }%
 }{%
 }%
\@ifundefined{endabstract}{\def\endabstract
  {\if@twocolumn\else\endquotation\fi}}{}%
\@ifundefined{maketitle}{\def\maketitle#1{}}{}%
\@ifundefined{affiliation}{\def\affiliation#1{}}{}%
\@ifundefined{proof}{}{}%
\@ifundefined{endproof}{}{}%
\@ifundefined{newfield}{\def\newfield#1#2{}}{}%
\@ifundefined{chapter}{\def\chapter#1{\par(Chapter head:)#1\par }%
 \newcount\c@chapter}{}%
\@ifundefined{part}{\def\part#1{\par(Part head:)#1\par }}{}%
\@ifundefined{section}{\def\section#1{\par(Section head:)#1\par }}{}%
\@ifundefined{subsection}{\def\subsection#1%
 {\par(Subsection head:)#1\par }}{}%
\@ifundefined{subsubsection}{\def\subsubsection#1%
 {\par(Subsubsection head:)#1\par }}{}%
\@ifundefined{paragraph}{\def\paragraph#1%
 {\par(Subsubsubsection head:)#1\par }}{}%
\@ifundefined{subparagraph}{\def\subparagraph#1%
 {\par(Subsubsubsubsection head:)#1\par }}{}%
\@ifundefined{therefore}{}{}%
\@ifundefined{backepsilon}{}{}%
\@ifundefined{yen}{}{}%
\@ifundefined{registered}{%
   \def\registered{\relax\ifmmode{}\r@gistered
                    \else$\m@th\r@gistered$\fi}%
 \def\r@gistered{^{\ooalign
  {\hfil\raise.07ex\hbox{$\scriptstyle\rm\text{R}$}\hfil\crcr
  \mathhexbox20D}}}}{}%
\@ifundefined{Eth}{}{}%
\@ifundefined{eth}{}{}%
\@ifundefined{Thorn}{}{}%
\@ifundefined{thorn}{}{}%
\@ifundefined{degree}{}{}%
\newdimen\theight
\def\Column{%
 \vadjust{\setbox\z@=\hbox{\scriptsize\quad\quad tcol}%
  \theight=\ht\z@\advance\theight by \dp\z@\advance\theight by \lineskip
  \kern -\theight \vbox to \theight{%
   \rightline{\rlap{\box\z@}}%
   \vss
   }%
  }%
 }%
\def\qed{%
 \ifhmode\unskip\nobreak\fi\ifmmode\ifinner\else\hskip5\p@\fi\fi
 \hbox{\hskip5\p@\vrule width4\p@ height6\p@ depth1.5\p@\hskip\p@}%
 }%
\def\miss{\hbox{\vrule height2\p@ width 2\p@ depth\z@}}%
\def\tcol#1{{\baselineskip=6\p@ \vcenter{#1}} \Column}  %
\def\newfmtname{LaTeX2e}
\def\chkcompat{%
   \if@compatibility
   \else
     \usepackage{latexsym}
   \fi
}

\ifx\fmtname\newfmtname
  \DeclareOldFontCommand{\rm}{\normalfont\rmfamily}{\mathrm}
  \DeclareOldFontCommand{\sf}{\normalfont\sffamily}{\mathsf}
  \DeclareOldFontCommand{\tt}{\normalfont\ttfamily}{\mathtt}
  \DeclareOldFontCommand{\bf}{\normalfont\bfseries}{\mathbf}
  \DeclareOldFontCommand{\it}{\normalfont\itshape}{\mathit}
  \DeclareOldFontCommand{\sl}{\normalfont\slshape}{\@nomath\sl}
  \DeclareOldFontCommand{\sc}{\normalfont\scshape}{\@nomath\sc}
  \chkcompat
\fi

\def\alpha{{\Greekmath 010B}}%
\def\beta{{\Greekmath 010C}}%
\def\gamma{{\Greekmath 010D}}%
\def\delta{{\Greekmath 010E}}%
\def\epsilon{{\Greekmath 010F}}%
\def\zeta{{\Greekmath 0110}}%
\def\eta{{\Greekmath 0111}}%
\def\theta{{\Greekmath 0112}}%
\def\iota{{\Greekmath 0113}}%
\def\kappa{{\Greekmath 0114}}%
\def\lambda{{\Greekmath 0115}}%
\def\mu{{\Greekmath 0116}}%
\def\nu{{\Greekmath 0117}}%
\def\xi{{\Greekmath 0118}}%
\def\pi{{\Greekmath 0119}}%
\def\rho{{\Greekmath 011A}}%
\def\sigma{{\Greekmath 011B}}%
\def\tau{{\Greekmath 011C}}%
\def\upsilon{{\Greekmath 011D}}%
\def\phi{{\Greekmath 011E}}%
\def\chi{{\Greekmath 011F}}%
\def\psi{{\Greekmath 0120}}%
\def\omega{{\Greekmath 0121}}%
\def\varepsilon{{\Greekmath 0122}}%
\def\vartheta{{\Greekmath 0123}}%
\def\varpi{{\Greekmath 0124}}%
\def\varrho{{\Greekmath 0125}}%
\def\varsigma{{\Greekmath 0126}}%
\def\varphi{{\Greekmath 0127}}%

\def\nabla{{\Greekmath 0272}}
\def\FindBoldGroup{%
   {\setbox0=\hbox{$\mathbf{x\global\edef\theboldgroup{\the\mathgroup}}$}}%
}

\def\Greekmath#1#2#3#4{%
    \if@compatibility
        \ifnum\mathgroup=\symbold
           \mathchoice{\mbox{\boldmath$\displaystyle\mathchar"#1#2#3#4$}}%
                      {\mbox{\boldmath$\textstyle\mathchar"#1#2#3#4$}}%
                      {\mbox{\boldmath$\scriptstyle\mathchar"#1#2#3#4$}}%
                      {\mbox{\boldmath$\scriptscriptstyle\mathchar"#1#2#3#4$}}%
        \else
           \mathchar"#1#2#3#4%
        \fi 
    \else 
        \FindBoldGroup
        \ifnum\mathgroup=\theboldgroup 
           \mathchoice{\mbox{\boldmath$\displaystyle\mathchar"#1#2#3#4$}}%
                      {\mbox{\boldmath$\textstyle\mathchar"#1#2#3#4$}}%
                      {\mbox{\boldmath$\scriptstyle\mathchar"#1#2#3#4$}}%
                      {\mbox{\boldmath$\scriptscriptstyle\mathchar"#1#2#3#4$}}%
        \else
           \mathchar"#1#2#3#4%
        \fi                 
          \fi}

\newif\ifGreekBold  \GreekBoldfalse
\let\SAVEPBF=\pbf
\def\pbf{\GreekBoldtrue\SAVEPBF}%

\@ifundefined{theorem}{}{}
\@ifundefined{lemma}{}{}
\@ifundefined{corollary}{}{}
\@ifundefined{conjecture}{}{}
\@ifundefined{proposition}{}{}
\@ifundefined{axiom}{}{}
\@ifundefined{remark}{}{}
\@ifundefined{example}{}{}
\@ifundefined{exercise}{}{}
\@ifundefined{definition}{}{}

\@ifundefined{mathletters}{%
  \newcounter{equationnumber}  
  \def\mathletters{%
     \addtocounter{equation}{1}
     \edef\@currentlabel{\theequation}%
     \setcounter{equationnumber}{\c@equation}
     \setcounter{equation}{0}%
     \edef\theequation{\@currentlabel\noexpand\alph{equation}}%
  }
  
}{}

\@ifundefined{BibTeX}{%
    \def\BibTeX{{\rm B\kern-.05em{\sc i\kern-.025em b}\kern-.08em
                 T\kern-.1667em\lower.7ex\hbox{E}\kern-.125emX}}}{}%
\@ifundefined{AmS}%
    {\def\AmS{{\protect\usefont{OMS}{cmsy}{m}{n}%
                A\kern-.1667em\lower.5ex\hbox{M}\kern-.125emS}}}{}%
\@ifundefined{AmSTeX}{}{}%
\ifx\ds@amstex\relax
\makeatother\endinput
\else
   \@ifpackageloaded{amstex}%
      {\message{amstex already loaded}\makeatother\endinput}
      {}
   \@ifpackageloaded{amsgen}%
      {\message{amsgen already loaded}\makeatother\endinput}
      {}
\fi
\let\DOTSI\relax
\def\RIfM@{\relax\ifmmode}%
\def\FN@{\futurelet\next}%
\newcount\intno@
\def\iint{\DOTSI\intno@\tw@\FN@\ints@}%
\def\iiint{\DOTSI\intno@\thr@@\FN@\ints@}%
\def\iiiint{\DOTSI\intno@4 \FN@\ints@}%
\def\idotsint{\DOTSI\intno@\z@\FN@\ints@}%
\def\ints@{\findlimits@\ints@@}%
\newif\iflimtoken@
\newif\iflimits@
\def\findlimits@{\limtoken@true\ifx\next\limits\limits@true
 \else\ifx\next\nolimits\limits@false\else
 \limtoken@false\ifx\ilimits@\nolimits\limits@false\else
 \ifinner\limits@false\else\limits@true\fi\fi\fi\fi}%
\def\multint@{\int\ifnum\intno@=\z@\intdots@                          
 \else\intkern@\fi                                                    
 \ifnum\intno@>\tw@\int\intkern@\fi                                   
 \ifnum\intno@>\thr@@\int\intkern@\fi                                 
 \int}
\def\multintlimits@{\intop\ifnum\intno@=\z@\intdots@\else\intkern@\fi
 \ifnum\intno@>\tw@\intop\intkern@\fi
 \ifnum\intno@>\thr@@\intop\intkern@\fi\intop}%
\def\intic@{%
    \mathchoice{\hskip.5em}{\hskip.4em}{\hskip.4em}{\hskip.4em}}%
\def\negintic@{\mathchoice
 {\hskip-.5em}{\hskip-.4em}{\hskip-.4em}{\hskip-.4em}}%
\def\ints@@{\iflimtoken@                                              
 \def\ints@@@{\iflimits@\negintic@
   \mathop{\intic@\multintlimits@}\limits                             
  \else\multint@\nolimits\fi                                          
  \eat@}
 \else                                                                
 \def\ints@@@{\iflimits@\negintic@
  \mathop{\intic@\multintlimits@}\limits\else
  \multint@\nolimits\fi}\fi\ints@@@}%
\def\intkern@{\mathchoice{\!\!\!}{\!\!}{\!\!}{\!\!}}%
\def\plaincdots@{\mathinner{\cdotp\cdotp\cdotp}}%
\def\intdots@{\mathchoice{\plaincdots@}%
 {{\cdotp}\mkern1.5mu{\cdotp}\mkern1.5mu{\cdotp}}%
 {{\cdotp}\mkern1mu{\cdotp}\mkern1mu{\cdotp}}%
 {{\cdotp}\mkern1mu{\cdotp}\mkern1mu{\cdotp}}}%
\def\RIfM@{\relax\protect\ifmmode}
\def\text{\RIfM@\expandafter\text@\else\expandafter\mbox\fi}
\let\nfss@text\text
\def\text@#1{\mathchoice
   {\textdef@\displaystyle\f@size{#1}}%
   {\textdef@\textstyle\tf@size{\firstchoice@false #1}}%
   {\textdef@\textstyle\sf@size{\firstchoice@false #1}}%
   {\textdef@\textstyle \ssf@size{\firstchoice@false #1}}%
   \glb@settings}

\def\textdef@#1#2#3{\hbox{{%
                    \everymath{#1}%
                    \let\f@size#2\selectfont
                    #3}}}
\newif\iffirstchoice@
\firstchoice@true
\def\Let@{\relax\iffalse{\fi\let\\=\cr\iffalse}\fi}%
\def\vspace@{\def\vspace##1{\crcr\noalign{\vskip##1\relax}}}%
\def\multilimits@{\bgroup\vspace@\Let@
 \baselineskip\fontdimen10 \scriptfont\tw@
 \advance\baselineskip\fontdimen12 \scriptfont\tw@
 \lineskip\thr@@\fontdimen8 \scriptfont\thr@@
 \lineskiplimit\lineskip
 \vbox\bgroup\ialign\bgroup\hfil$\m@th\scriptstyle{##}$\hfil\crcr}%
\def\Sb{_\multilimits@}%
\def\endSb{\crcr\egroup\egroup\egroup}%
\def\Sp{^\multilimits@}%

\newdimen\ex@
\def\rightarrowfill@#1{$#1\m@th\mathord-\mkern-6mu\cleaders
 \hbox{$#1\mkern-2mu\mathord-\mkern-2mu$}\hfill
 \mkern-6mu\mathord\to$}%
\def\leftarrowfill@#1{$#1\m@th\mathord\gets\mkern-6mu\cleaders
 \hbox{$#1\mkern-2mu\mathord-\mkern-2mu$}\hfill\mkern-6mu\mathord-$}%
\def\leftrightarrowfill@#1{$#1\m@th\mathord\gets
\mkern-6mu\cleaders
 \hbox{$#1\mkern-2mu\mathord-\mkern-2mu$}\hfill
 \mkern-6mu\mathord\to$}%
\def\overrightarrow{\mathpalette\overrightarrow@}%
\def\overrightarrow@#1#2{\vbox{\ialign{##\crcr\rightarrowfill@#1\crcr
 \noalign{\kern-\ex@\nointerlineskip}$\m@th\hfil#1#2\hfil$\crcr}}}%

\def\overleftarrow{\mathpalette\overleftarrow@}%
\def\overleftarrow@#1#2{\vbox{\ialign{##\crcr\leftarrowfill@#1\crcr
 \noalign{\kern-\ex@\nointerlineskip}$\m@th\hfil#1#2\hfil$\crcr}}}%
\def\overleftrightarrow{\mathpalette\overleftrightarrow@}%
\def\overleftrightarrow@#1#2{\vbox{\ialign{##\crcr
   \leftrightarrowfill@#1\crcr
 \noalign{\kern-\ex@\nointerlineskip}$\m@th\hfil#1#2\hfil$\crcr}}}%
\def\underrightarrow{\mathpalette\underrightarrow@}%
\def\underrightarrow@#1#2{\vtop{\ialign{##\crcr$\m@th\hfil#1#2\hfil
  $\crcr\noalign{\nointerlineskip}\rightarrowfill@#1\crcr}}}%

\def\underleftarrow{\mathpalette\underleftarrow@}%
\def\underleftarrow@#1#2{\vtop{\ialign{##\crcr$\m@th\hfil#1#2\hfil
  $\crcr\noalign{\nointerlineskip}\leftarrowfill@#1\crcr}}}%
\def\underleftrightarrow{\mathpalette\underleftrightarrow@}%
\def\underleftrightarrow@#1#2{\vtop{\ialign{##\crcr$\m@th
  \hfil#1#2\hfil$\crcr
 \noalign{\nointerlineskip}\leftrightarrowfill@#1\crcr}}}%

\def\qopnamewl@#1{\mathop{\operator@font#1}\nlimits@}
\let\nlimits@\displaylimits
\def\setboxz@h{\setbox\z@\hbox}

\def\varlim@#1#2{\mathop{\vtop{\ialign{##\crcr
 \hfil$#1\m@th\operator@font lim$\hfil\crcr
 \noalign{\nointerlineskip}#2#1\crcr
 \noalign{\nointerlineskip\kern-\ex@}\crcr}}}}

 \def\rightarrowfill@#1{\m@th\setboxz@h{$#1-$}\ht\z@\z@
  $#1\copy\z@\mkern-6mu\cleaders
  \hbox{$#1\mkern-2mu\box\z@\mkern-2mu$}\hfill
  \mkern-6mu\mathord\to$}
\def\leftarrowfill@#1{\m@th\setboxz@h{$#1-$}\ht\z@\z@
  $#1\mathord\gets\mkern-6mu\cleaders
  \hbox{$#1\mkern-2mu\copy\z@\mkern-2mu$}\hfill
  \mkern-6mu\box\z@$}

\def\projlim{\qopnamewl@{proj\,lim}}
\def\injlim{\qopnamewl@{inj\,lim}}
\def\varinjlim{\mathpalette\varlim@\rightarrowfill@}
\def\varprojlim{\mathpalette\varlim@\leftarrowfill@}
\def\varliminf{\mathpalette\varliminf@{}}
\def\varliminf@#1{\mathop{\underline{\vrule\@depth.2\ex@\@width\z@
   \hbox{$#1\m@th\operator@font lim$}}}}
\def\varlimsup{\mathpalette\varlimsup@{}}
\def\varlimsup@#1{\mathop{\overline
  {\hbox{$#1\m@th\operator@font lim$}}}}

\begingroup \catcode `|=0 \catcode `[= 1
\catcode`]=2 \catcode `\{=12 \catcode `\}=12
\catcode`\\=12 
|gdef|@alignverbatim#1\end{align}[#1|end[align]]
|gdef|@salignverbatim#1\end{align*}[#1|end[align*]]

|gdef|@alignatverbatim#1\end{alignat}[#1|end[alignat]]
|gdef|@salignatverbatim#1\end{alignat*}[#1|end[alignat*]]

|gdef|@xalignatverbatim#1\end{xalignat}[#1|end[xalignat]]
|gdef|@sxalignatverbatim#1\end{xalignat*}[#1|end[xalignat*]]

|gdef|@gatherverbatim#1\end{gather}[#1|end[gather]]
|gdef|@sgatherverbatim#1\end{gather*}[#1|end[gather*]]

|gdef|@gatherverbatim#1\end{gather}[#1|end[gather]]
|gdef|@sgatherverbatim#1\end{gather*}[#1|end[gather*]]

|gdef|@multilineverbatim#1\end{multiline}[#1|end[multiline]]
|gdef|@smultilineverbatim#1\end{multiline*}[#1|end[multiline*]]

|gdef|@arraxverbatim#1\end{arrax}[#1|end[arrax]]
|gdef|@sarraxverbatim#1\end{arrax*}[#1|end[arrax*]]

|gdef|@tabulaxverbatim#1\end{tabulax}[#1|end[tabulax]]
|gdef|@stabulaxverbatim#1\end{tabulax*}[#1|end[tabulax*]]

|endgroup

\def\align{\@verbatim \frenchspacing\@vobeyspaces \@alignverbatim
You are using the "align" environment in a style in which it is not defined.}

\@namedef{align*}{\@verbatim\@salignverbatim
You are using the "align*" environment in a style in which it is not defined.}
\expandafter\let\csname endalign*\endcsname =\endtrivlist

\def\alignat{\@verbatim \frenchspacing\@vobeyspaces \@alignatverbatim
You are using the "alignat" environment in a style in which it is not defined.}

\@namedef{alignat*}{\@verbatim\@salignatverbatim
You are using the "alignat*" environment in a style in which it is not defined.}
\expandafter\let\csname endalignat*\endcsname =\endtrivlist

\def\xalignat{\@verbatim \frenchspacing\@vobeyspaces \@xalignatverbatim
You are using the "xalignat" environment in a style in which it is not defined.}

\@namedef{xalignat*}{\@verbatim\@sxalignatverbatim
You are using the "xalignat*" environment in a style in which it is not defined.}
\expandafter\let\csname endxalignat*\endcsname =\endtrivlist

\def\gather{\@verbatim \frenchspacing\@vobeyspaces \@gatherverbatim
You are using the "gather" environment in a style in which it is not defined.}

\@namedef{gather*}{\@verbatim\@sgatherverbatim
You are using the "gather*" environment in a style in which it is not defined.}
\expandafter\let\csname endgather*\endcsname =\endtrivlist

\def\multiline{\@verbatim \frenchspacing\@vobeyspaces \@multilineverbatim
You are using the "multiline" environment in a style in which it is not defined.}

\@namedef{multiline*}{\@verbatim\@smultilineverbatim
You are using the "multiline*" environment in a style in which it is not defined.}
\expandafter\let\csname endmultiline*\endcsname =\endtrivlist

\def\arrax{\@verbatim \frenchspacing\@vobeyspaces \@arraxverbatim
You are using a type of "array" construct that is only allowed in AmS-LaTeX.}

\def\tabulax{\@verbatim \frenchspacing\@vobeyspaces \@tabulaxverbatim
You are using a type of "tabular" construct that is only allowed in AmS-LaTeX.}

\@namedef{arrax*}{\@verbatim\@sarraxverbatim
You are using a type of "array*" construct that is only allowed in AmS-LaTeX.}
\expandafter\let\csname endarrax*\endcsname =\endtrivlist

\@namedef{tabulax*}{\@verbatim\@stabulaxverbatim
You are using a type of "tabular*" construct that is only allowed in AmS-LaTeX.}
\expandafter\let\csname endtabulax*\endcsname =\endtrivlist

\def\@@eqncr{\let\@tempa\relax
    \ifcase\@eqcnt \def\@tempa{& & &}\or \def\@tempa{& &}%
      \else \def\@tempa{&}\fi
     \@tempa
     \if@eqnsw
        \iftag@
           \@taggnum
        \else
           \@eqnnum\stepcounter{equation}%
        \fi
     \fi
     \global\tag@false
     \global\@eqnswtrue
     \global\@eqcnt\z@\cr}

 \def\endequation{%
     \ifmmode\ifinner 
      \iftag@
        \addtocounter{equation}{-1} 
        $\hfil
           \displaywidth\linewidth\@taggnum\egroup \endtrivlist
        \global\tag@false
        \global\@ignoretrue   
      \else
        $\hfil
           \displaywidth\linewidth\@eqnnum\egroup \endtrivlist
        \global\tag@false
        \global\@ignoretrue 
      \fi
     \else   
      \iftag@
        \addtocounter{equation}{-1} 
        \eqno \hbox{\@taggnum}
        \global\tag@false%
        $$\global\@ignoretrue
      \else
        \eqno \hbox{\@eqnnum}
        $$\global\@ignoretrue
      \fi
     \fi\fi
 } 

 \newif\iftag@ \tag@false
 
 \def\tag{\@ifnextchar*{\@tagstar}{\@tag}}
 \def\@tag#1{%
     \global\tag@true
     \global\def\@taggnum{(#1)}}
 \def\@tagstar*#1{%
     \global\tag@true
     \global\def\@taggnum{#1}%
}

\makeatother

\begin{document}
\title{\normalsize
\qquad\qquad\qquad\qquad\qquad\qquad\qquad\qquad\qquad\qquad\qquad\qquad\qquad\qquad\qquad
\vspace{20mm}
UCL-IPT-99-09
\LARGE
Gauge freedom for Gravitational Wave problems in tensor-scalar theories of
gravity\vspace{1cm}}
\author{
Yves Wiaux\\
\\
Institut de Physique Th\'{e}orique\\
Universit\'{e} catholique de Louvain\\
B-1348 Louvain-la-Neuve, Belgium}
\date{August 1999}
\maketitle
\vspace{10mm}


\begin{abstract}
A specific choice of gauge is shown to imply a decoupling between the tensor and
scalar components of Gravitational Radiation in the context of Brans-Dicke
type theories of gravitation. The comparison of the predictions of these
theories with those of General Relativity is thereby made straightforward.
\end{abstract}

\begin{center}
\pagebreak
\end{center}

\section{Introduction}

The fundamental postulate at the basis of any theory of gravitation is that
of diffeomorphism invariance under general coordinate transformations. As a
consequence, physically equivalent solutions to the ensuing dynamics fall
into gauge equivalence classes, namely gauge orbits under the diffeomorphism
group.

In General Relativity, the dynamics of the spacetime metric is solely
determined from Einstein's equations. As just pointed out, a unique physical
solution, associated to a unique geometry of the Universe, may be
represented in terms of different functional solutions for the metric field,
each corresponding to a different coordinate system, chosen to parametrize
this geometry. Just like for Maxwell's equations for electromagnetism, any
theory for gravity thus enjoys such a gauge freedom, which can be used to
simplify the field equations and facilitate the interpretation of any
phenomenon predicted by the theory.\ In particular, there exists some
appropriate choice of gauge fixing when tackling the problem of
Gravitational Waves.

Beyond General Relativity, a scalar gravitational sector has been largely
suggested by the most promising theories for the description of fundamental
interactions, that is, String Theories \cite{green}. Among all tensor-scalar
models, Brans-Dicke theories correspond to the simplest generalization of
Einstein's theory including a scalar gravitational component. Obviously, the
symmetry under general coordinate transformations remains valid in all
consistent theories beyond General Relativity.

The aim of this letter is to discuss a choice of gauge fixing which seems to
be particularly appropriate when tackling the problem of Gravitational Waves
in tensor-scalar theories. But first, we briefly review the conventional
gauge fixing choices in General Relativity and Brans-Dicke type theories.

\section{Within the limits of General Relativity}

In a weak field region, we can expand the metric field (in cartesian
coordinates) in terms of a small perturbation $h_{\mu \nu }$ around
Minkowski's metric $\eta _{\mu \nu }$ \footnote{%
Our signature convention for $\eta _{\mu \nu }$ is $\left( +,-,-,-\right) $.%
}, namely 
\begin{equation}
g_{\mu \nu }=\eta _{\mu \nu }+\sqrt{2\kappa}h_{\mu \nu }\text{ ,}  \label{1}
\end{equation}
where 
\begin{equation}
\kappa =\frac{8\pi G}{c^{4}}\text{ .}  \label{2}
\end{equation}
The linearized Einstein field equations for $h_{\mu \nu }$ then read in any
frame (indices being raised by means of $\eta _{\mu \nu }$ ) \cite{weinberg}%
: 
\begin{equation}
\Box h_{\mu \nu }-h_{\hspace{0.13cm}\nu ,\mu ,\lambda }^{\lambda }+h_{%
\hspace{0.13cm}\lambda ,\mu ,\nu }^{\lambda }-h_{\hspace{0.13cm}\mu ,\nu
,\lambda }^{\lambda }=-\sqrt{2\kappa }S_{\mu \nu }\text{ ,}  \label{3}
\end{equation}
with 
\begin{equation}
S_{\mu \nu }=T_{\mu \nu }-\frac{1}{2}\eta _{\mu \nu }T\text{ , }
T=T_{\hspace{0.13cm}\lambda }^{\lambda }\text{ .}  \label{4}
\end{equation}
$T_{\mu \nu }$ being the conserved stress-energy tensor for matter fields,
computed on Minkowski's spacetime.

Consider then an infinitesimal coordinate transformation 
\begin{equation}
x^{\prime \mu }=x^{\mu }+\varepsilon ^{\mu }\left( x\right) \text{ ,}
\label{5}
\end{equation}
with $\varepsilon ^{\mu }\left( x\right) $ of the same order of magnitude as
the perturbation of the metric field. The metric being a second rank tensor,
to first order the following transformation law for $h_{\mu \nu }$ then
applies, 
\begin{equation}
\sqrt{2\kappa }h_{\mu \nu }^{\prime }=\sqrt{2\kappa }h_{\mu \nu }-\partial
_{\left( \mu \right. }\varepsilon _{\left. \nu \right) }\text{ ,}  \label{6}
\end{equation}
which means that, if $h_{\mu \nu }$ is a solution to the equations (\ref{3})
in the original frame, then $h_{\mu \nu }^{\prime }$, defined by (\ref{6}), is
another expression for the same physical solution in the frame obtained
through the transformation (\ref{5}).

The analysis of Gravitational Waves is always discussed in a class of
''Harmonic'' coordinate systems \cite{weinberg}, defined by the four
following conditions: 
\begin{equation}
\partial _{\mu }\bar{h}^{\mu \nu }=0\text{ ,}  \label{7}
\end{equation}
with 
\begin{equation}
\stackrel{\_}{h}_{\mu \nu }\equiv h_{\mu \nu }-\frac{1}{2}\eta _{\mu \nu }h,%
\text{\ }h=h_{\hspace{0.13cm}\lambda }^{\lambda }\text{ .}  \label{8}
\end{equation}
These restrictions provide a simplified expression for the field equations 
(\ref{3}): 
\begin{equation}
\Box \bar{h}_{\mu \nu }=-\sqrt{2\kappa }T_{\mu \nu }\text{ ,}  \label{9}
\end{equation}
a clear indication for the existence of propagating wavelike solutions for
the perturbation $h_{\mu \nu }$. However, a residual gauge freedom of the
form (\ref{5}) remains within this class of frames, constrained by the
conditions 
\begin{equation}
\Box \varepsilon ^{\mu }\left( x\right) =0\text{ ,}  \label{10}
\end{equation}
whose general solution is thus of the form 
\begin{equation}
\varepsilon ^{\mu }\left( x\right) =i\varepsilon ^{\mu }\left( k\right)
e^{ik^{\lambda }x_{\lambda }}\text{ .}  \label{11}
\end{equation}

The choice of a specific frame thus requires 4 additional constraints,
bringing the number of independent polarization states of the waves from 10 (%
$h_{\mu \nu }$ being a symmetric tensor) down to 2. To understand the impact
of a Gravitational Wave impinging on a system of masses, it is far easier to
consider this process in a comobile frame, that is, a coordinate system in
which the grid of coordinates moves with the bodies \cite{schutz}. In such a
frame, one can express the gauge-invariant element of proper distance
between two objects as 
\begin{equation}
d\sigma ^{2}=\left( -\delta _{ij}+h_{ij}\right) dx^{i}dx^{j}\text{ ,}
\label{12}
\end{equation}
where the coordinate difference $dx^{i}$ is constant during the impact of
the wave. One can thus easily understand how this distance gets modified by
a Gravitational Wave.

These points having been recalled, let us now indicate why one usually
speaks of a Transverse-Traceless (TT) gauge: the geodesic equation for a
particle initially at rest reads, at $t=0$ 
\begin{equation}
\frac{d^{2}x^{i}}{dt^{2}}=\frac{1}{2}\left( \frac{cdt}{d\tau }\right)
^{2}\left( 2h_{\hspace{0.13cm}0,0}^{i}-h_{00}^{\hspace{0.26cm},i}\right) 
\text{ .}  \label{13}
\end{equation}
Hence, in order to keep fixed coordinate differences between bodies 
\begin{equation}
\frac{dx^{i}}{dt}=0\text{ ,}  \label{13a}
\end{equation}
we need to restrict to configurations such that 
\begin{equation}
h_{0\mu }=cst.  \label{14}
\end{equation}
It is readily shown that the 8 conditions (\ref{7}) and (\ref{14}) are also
implemented by 
\begin{equation}
\partial _{i}h^{ij}=0\text{ ,}  \label{15}
\end{equation}
\begin{equation}
h=0  \label{16}
\end{equation}
and 
\begin{equation}
h_{0\mu }=0\text{ ,}  \label{17}
\end{equation}
that is, by virtue of (\ref{15}) and (\ref{16}), the waves are \emph{transverse}
and \emph{traceless} in a comobile frame. This justifies the name ''TT-gauge''.
Note that the physical effect of a Gravitational Wave on the proper distance
(\ref{12}) is actually transverse and traceless, as it may be readily seen by
working in a comobile frame.

The point is that the interpretation of the wave impact would not be so
direct in any other frame, where the geodesic equation cannot be reduced to 
(\ref{13a}). But, if one considers the existence of Gravitational Waves in a
tensor-scalar theory, other considerations may lead to a different choice of
gauge. A discussion of this specific issue is the purpose of the following
section.

\section{The case of Brans-Dicke theories}

The action for Brans-Dicke theories reads: 
\begin{equation}
S=-\frac{1}{16\pi }\int d^{4}x\sqrt{g}\left( \Phi R-\frac{\omega_{BD} }{\Phi }%
\Phi _{,\mu }\Phi ^{,\mu }\right) +\int d^{4}x{\pounds}_{mat} \left( \Psi ,g_{\mu
\nu }\right)  \label{b1}
\end{equation}
where $\Phi $ is the scalar gravitational component and $\Psi $ stands for
matter fields. For simplicity, we define the tensor and scalar
perturbations, respectively on Minkowski's spacetime and around a constant
expectation value for the scalar field: 
\begin{equation}
g_{\mu \nu }=\eta _{\mu \nu }+\sqrt{2\bar{\kappa}}h_{\mu \nu }\text{ ,}
\label{b2}
\end{equation}
and 
\begin{equation}
\Phi =\Phi _{0}\left( 1+a_{BD}\sqrt{2\bar{\kappa}}\varphi \right) \text{ ,}
\label{b3}
\end{equation}
where 
\begin{equation}
\bar{\kappa}=\frac{8\pi }{\Phi _{0}c^{4}}  \label{b4}
\end{equation}
stands for the gravitational coupling, analogous to $\kappa $ in General
Relativity, while $a_{BD}$ represents the scalar coupling to matter fields,
defined as $a_{BD}^{2}=1/(2 \omega_{BD} + 3)$ .
The linearization of the field equations for $h_{\mu \nu }$ and $\varphi $ 
\cite{fucito} then implies 
\begin{equation}
\begin{tabular}{r}
$-\frac{1}{2}\left[ \Box h_{\mu \nu }-h_{\hspace{0.13cm}\nu ,\mu ,\lambda
}^{\lambda }+h_{\hspace{0.13cm}\lambda ,\mu ,\nu }^{\lambda }-h_{\hspace{%
0.13cm}\mu ,\nu ,\lambda }^{\lambda }\right] -\frac{1}{2}\eta _{\mu \nu
}\left[ -\Box h+h_{\hspace{0.26cm},\alpha ,\beta }^{\alpha \beta }\right] -$
\\ 
$\left[ a_{BD}\varphi _{,\mu ,\nu }-a_{BD}\varphi _{\hspace{0.13cm},\lambda
}^{,\lambda }\eta _{\mu \nu }\right] =\sqrt{\frac{\bar{\kappa}}{2}}T_{\mu \nu }$%
\end{tabular}
\end{equation}
and 
\begin{equation}
\Box \varphi =a_{BD}\sqrt{\frac{\bar{\kappa}}{2}}T%
\text{ .}  \label{b6}
\end{equation}
Invariance under general coordinate transformations results once again in
the transformation law (\ref{6}) for the metric perturbation, while the scalar
field remains invariant to first order.

Defining a symmetric tensor field \cite{fucito} 
\begin{equation}
\Theta _{\mu \nu }\equiv h_{\mu \nu }-\frac{1}{2}\eta _{\mu \nu
}h-a_{BD}\eta _{\mu \nu }\varphi \text{ }\equiv \bar{h}_{\mu \nu
}-a_{BD}\eta _{\mu \nu }\varphi  \label{b7}
\end{equation}
which thus transforms according to 
\begin{equation}
\sqrt{2\bar{\kappa}}\Theta _{\mu \nu }^{\prime }=\sqrt{2\bar{\kappa}}\Theta
_{\mu \nu }-\partial _{\left( \mu \right. }\varepsilon _{\left. \nu \right)
}+\eta _{\mu \nu }\partial ^{\lambda }\varepsilon _{\lambda }%
\text{ ,}  \label{b7a}
\end{equation}
one can impose the following gauge conditions 
\begin{equation}
\partial ^{\mu }\Theta _{\mu \nu }=0\text{ .}  \label{b8}
\end{equation}
Once more, the field equations then simplify to ordinary wave equations,
namely 
\begin{equation}
-\frac{1}{2}\Box \Theta _{\mu \nu }=\sqrt{\frac{\bar{\kappa}}{2}}T_{\mu \nu }
\label{b9}
\end{equation}
as well as 
\begin{equation}
\Box \varphi =a_{BD}\sqrt{\frac{\bar{\kappa}}{2}}T\text{ .}  \label{b10}
\end{equation}
The same residual conditions remain as specified in (\ref{10}), associated to
a large residual choice of gauge conditions. Any particular choice leaves
independent 3 (of the 11 initial) components of the extended gravitational
field. The following paragraphs are devoted to the interest of two specific
choices, one of them being, to our knowledge, rather unconventional, but
nevertheless extremely interesting.

The standard reference corresponds to the following set of conditions: 
\begin{equation}
\partial ^{\mu }\Theta _{\mu \nu }=0,\partial ^{i}\Theta _{ij}=0\text{ and }%
\Theta _{\mu \nu }=h_{\mu \nu }\text{ .}  \label{b11}
\end{equation}
Even though the conditions 
\begin{equation}
\Theta _{\mu \nu }=h_{\mu \nu }  \label{b12}
\end{equation}
seem natural in order to avoid a multiplication in the number of variables 
\cite{fucito}, they are not justified by themselves. These conditions are in
fact equivalent to 
\begin{equation}
\partial ^{\mu }\Theta _{\mu \nu }=0,\;\Theta _{\mu 0}=cst.\text{\ and\ }%
\Theta _{\mu \nu }=h_{\mu \nu }\text{ ,}  \label{b13}
\end{equation}
which imply in particular the constraints 
\begin{equation}
h_{\mu 0}=cst.\text{ ,}  \label{b14}
\end{equation}
and thereby define a comobile frame, whose usefulness need not be
emphasized anymore. Whatever the gauge, one can easily understand that 2 of
the 3 polarization states are helicity-2 ones (h=2), while the remaining one
is scalar (h=0). However, the precise form of each polarization determines a
specific perturbation in the proper distance between two masses in this
comobile frame only. The general expression for the wave then reads (for a
wave travelling in the z-direction): 
\begin{equation}
\begin{tabular}{l}
$\Theta _{\mu \nu }\left( x\right) =h_{\mu \nu }\left( x\right) =$ \\ 
$\vspace{0.3cm}$ \\ 
$h_{+}\left( x\right) \left[ 
\begin{array}{cccc}
0 & 0 & 0 & 0 \\ 
0 & 1 & 0 & 0 \\ 
0 & 0 & -1 & 0 \\ 
0 & 0 & 0 & 0
\end{array}
\right] +h_{\times }\left( x\right) \left[ 
\begin{array}{cccc}
0 & 0 & 0 & 0 \\ 
0 & 0 & 1 & 0 \\ 
0 & 1 & 0 & 0 \\ 
0 & 0 & 0 & 0
\end{array}
\right] +h_{scal}\left( x\right) \left[ 
\begin{array}{cccc}
0 & 0 & 0 & 0 \\ 
0 & -1 & 0 & 0 \\ 
0 & 0 & -1 & 0 \\ 
0 & 0 & 0 & 0
\end{array}
\right] $%
\end{tabular}
\label{b14a}
\end{equation}
where $h_{+}\left( x\right) $, $h_{\times }\left( x\right) $ and $%
h_{scal}\left( x\right) $ are the waveforms associated to the three
different polarization states.

Let us now consider the problem from a different point of view: the presence
of a scalar polarization of the waves, coupled to ordinary tensor waves, may
complicate somewhat the calculation leading to any physical result related
to the emission of gravitational radiation. One can thus consider, as a
criterium for gauge fixing, \textbf{the possibility of decoupling the scalar
and tensor parts }of the problem. This criterium may indeed be met by
imposing another set of conditions: 
\begin{equation}
\partial ^{\mu }\Theta _{\mu \nu }=0,\;\partial ^{i}\Theta _{ij}=0\text{\
and\ }\Theta =0\text{ .}  \label{b15}
\end{equation}
This choice of frame defines a gauge in which the polarization states of $%
\Theta _{\mu \nu }$ are transverse and traceless (call it, by analogy, the $%
\Theta $TT-gauge). For a wave travelling in the z-direction, the general
form for $\Theta _{\mu \nu }$ reads: 
\begin{equation}
\Theta _{\mu \nu }\left( x\right) =\Theta _{+}\left( x\right) \left[ 
\begin{array}{cccc}
0 & 0 & 0 & 0 \\ 
0 & 1 & 0 & 0 \\ 
0 & 0 & -1 & 0 \\ 
0 & 0 & 0 & 0
\end{array}
\right] +\Theta _{\times }\left( x\right) \left[ 
\begin{array}{cccc}
0 & 0 & 0 & 0 \\ 
0 & 0 & 1 & 0 \\ 
0 & 1 & 0 & 0 \\ 
0 & 0 & 0 & 0
\end{array}
\right]  \label{b15aa}
\end{equation}
where $\Theta _{+}\left( x\right) $ and $\Theta _{\times }\left( x\right) $
are the amplitudes associated to the two polarization states of $\Theta
_{\mu \nu }$. Both helicity-2 polarization states are totally independent of
the scalar gravitational field. The variables $\Theta _{\mu \nu }$ (h=2) and $%
\varphi $ (h=0) have thus been rendered independent. Note that $\Theta _{\mu
\nu }\left( x\right) $ is now different from $h_{\mu \nu }\left( x\right) $;
the metric perturbation is still defined in terms of 3 polarization states:
by virtue of the definition (\ref{b7}) for $\Theta _{\mu \nu }$ and the
condition $\Theta =0$, we have indeed 
\begin{equation}
h_{\mu \nu }=\Theta _{\mu \nu }-\frac{1}{2}\eta _{\mu \nu }\Theta
-a_{BD}\eta _{\mu \nu }\varphi =\Theta _{\mu \nu }-a_{BD}\eta _{\mu \nu
}\varphi  \label{b15a}
\end{equation}
which, in this frame, gives 
\begin{equation}
\begin{tabular}{l}
$h_{\mu \nu }\left( x\right) =$ \\ 
$\vspace{0.3cm}$ \\ 
$h_{+}\left( x\right) \left[ 
\begin{array}{cccc}
0 & 0 & 0 & 0 \\ 
0 & 1 & 0 & 0 \\ 
0 & 0 & -1 & 0 \\ 
0 & 0 & 0 & 0
\end{array}
\right] +h_{\times }\left( x\right) \left[ 
\begin{array}{cccc}
0 & 0 & 0 & 0 \\ 
0 & 0 & 1 & 0 \\ 
0 & 1 & 0 & 0 \\ 
0 & 0 & 0 & 0
\end{array}
\right] +h_{scal}\left( x\right) \left[ 
\begin{array}{cccc}
1 & 0 & 0 & 0 \\ 
0 & -1 & 0 & 0 \\ 
0 & 0 & -1 & 0 \\ 
0 & 0 & 0 & -1
\end{array}
\right] $ .
\end{tabular}
\label{b16}
\end{equation}
Obviously, two helicity-2 polarizations remain, while the third one is still
scalar, but provides a conformally flat geometry to spacetime, which after
all is natural when considering a massless scalar perturbation!

However, the great interest of this $\Theta $TT-gauge \footnote{%
Note that in a recent paper \cite{maggiore} received after the present work was
completed, the same choice of gauge fixing is being discussed. However, these authors
do not seem to realize that the main interest of this
choice resides in the decoupling of the tensor and scalar components.} is
that, working with \textbf{independent variables}, it is possible to
reconstruct a simple effective theory for the study of tensor-scalar waves.
The corresponding action (the variation of which gives (\ref{b9}) and (\ref{b10})%
) does indeed separate into two parts: 
\begin{equation}
S=S_{2}+S_{0}  \label{b17}
\end{equation}
with 
\begin{equation}
S_{2}=\int d^{4}x\left[ \left( \frac{1}{4}\partial _{\alpha }\Theta ^{\mu
\nu }\partial ^{\alpha }\Theta _{\mu \nu }-\sqrt{\frac{\bar{\kappa}}{2}}\Theta ^{\mu
\nu }T_{\mu \nu }\right) +\lambda ^{\nu }\partial ^{\mu }\Theta _{\mu \nu
}+\lambda\Theta\right]   \label{b18}
\end{equation}
and 
\begin{equation}
S_{0}=\int d^{4}x\left[ \frac{1}{2}\partial _{\alpha }\varphi \partial
^{\alpha }\varphi +a_{BD}\sqrt{\frac{\bar{\kappa}}{2}}\varphi T\right] \text{ ,}
\label{b19}
\end{equation}
Where $\lambda ^{\nu }$ and $\lambda$ are Lagrange multipliers necessary to impose
the conditions $\partial ^{\mu }\Theta _{\mu \nu }=0$ and $\Theta=0$ only under
which the equations (\ref{b9}) hold.

This new formalism should allow for a simplified analysis of Gravitational
Wave phenomena within the context of tensor-scalar theories by making
straightforward the transposition of theoretical results from General
Relativity. By having decoupled the contributions of the tensor and scalar
components of gravitational radiation, assessing the impact of the scalar
gravitational component on physical observations should become feasable in a
much more transparent and direct way.

Even though further study should provide more examples, we only consider, in
the two last sections, the well-known problem of the energy emission rate by a
gravitational source in the quadrupolar approximation. We eventually apply
it to a determination of the scalar to tensor ratio for the energy emission
rate in the case of the Hulse-Taylor Binary Pulsar (PSR1913+16).

\section{Application...}

Using the new effective action, a semi-classical calculation \footnote{%
This calculation is performed following the same scheme as the one used in ref.
\cite{mohanty} within the specific context of General Relativity only, while we extend
it independently to both tensor and scalar emissions as predicted in the context
of Brans-Dicke theories.} for a quantized wave emitted by a classical
source, immediately leads to mean emission rates for tensor and scalar
radiation given by, respectively 
\begin{equation}
\left\langle -\frac{dE_{\left( h=2\right) }}{dt}\right\rangle =\frac{1}{%
5c^{5}\Phi _{0}}\left\langle \frac{d^{3}}{dt^{3}}Q^{ij*}\left( t\right) 
\frac{d^{3}}{dt^{3}}Q_{ij}\left( t\right) -\frac{1}{3}\frac{d^{3}}{dt^{3}}%
Q_{i}^{i*}\left( t\right) \frac{d^{3}}{dt^{3}}Q_{j}^{j}\left( t\right)
\right\rangle   \label{b20}
\end{equation}
and 
\begin{equation}
\left\langle -\frac{dE_{\left( h=0\right) }}{dt}\right\rangle =\frac{a_{BD}^{2}}{2}%
\frac{1}{5c^{5}\Phi _{0}}\left\langle \frac{d^{3}}{dt^{3}}Q_{i}^{i*}\left(
t\right) \frac{d^{3}}{dt^{3}}Q_{j}^{j}\left( t\right) +\frac{1}{3}\frac{d^{3}%
}{dt^{3}}Q^{ij*}\left( t\right) \frac{d^{3}}{dt^{3}}Q_{ij}\left( t\right)
\right\rangle \text{ ,}  \label{b21}
\end{equation}
where the $Q_{ij}\left( t\right) $ are the quadrupolar moments of the source 
\begin{equation}
Q_{ij}\left( t\right) \equiv \frac{1}{c^{2}}\int d^{3}\vec{x}^{\prime
}T_{00}\left( \vec{x}^{\prime },t\right) x_{i}^{\prime }x_{j}^{\prime
}\equiv \int d^{3}\vec{x}^{\prime }\rho \left( \vec{x}^{\prime },t\right)
x_{i}^{\prime }x_{j}^{\prime }\text{ ,}  \label{b22}
\end{equation}
while $\rho \left( \vec{x}^{\prime },t\right) $ is its matter density.

This clearly illustrates how theoretical results are straightforwardly extended
beyond those of General Relativity. The helicity-2 energy emission rate given in
 (\ref{b20}) is totally equivalent to what General Relativity predicts \footnote{%
With the remaining difference that the gravitational coupling is different from the
one of general relativity: $\kappa \not= \bar{\kappa}$.

One may thus define the mean expectation value for the scalar field $\Phi _{0}$ to be
$\Phi _{0}=\frac{1}{G}$, keeping in mind that G differs from Newton's constant.}, while
the additional contribution, given by (\ref{b21}), is shown to appear
as a consequence of the existence of the scalar component of gravitation in
Brans-Dicke theories.

\section{...To the Binary Pulsar PSR1913+16}

We may thus calculate now the correction induced by the scalar component of
gravitation to the acceleration of the orbital motion for the Hulse-Taylor
pulsar around its companion, this binary being, presently, our best laboratory for
the study of Gravitational Waves in the radiative-strong-field regime. A somewhat
long but direct enough calculation (extending the one done in ref. \cite{peters}
for the case of General Relativity, i.e., for the tensor part) gives, respectively, 
\begin{equation}
\left\langle -\frac{dE_{\left( h=2\right) }}{dt}\right\rangle =\frac{32}{5}%
\frac{G^{4}M^{3}\mu ^{2}}{c^{5}a^{5}}f\left( e\right)   \label{b23}
\end{equation}
and 
\begin{equation}
\left\langle -\frac{dE_{\left( h=0\right) }}{dt}\right\rangle =\left( \frac{%
a_{BD}^{2}}{6}\right) \frac{32}{5}\frac{G^{4}M^{3}\mu ^{2}}{c^{5}a^{5}}%
g\left( e\right) \text{ ,}  \label{b24}
\end{equation}
where $f\left( e\right) $ and $g\left( e\right) $ are enhancement factors,
relatively to the case of a circular orbit ($e$ being the eccentricity of
the orbit and $a$ its semi-major axis), defined by 
\begin{equation}
f\left( e\right) =\frac{\left( 1+\frac{73}{24}e^{2}+\frac{37}{96}%
e^{4}\right) }{\left( 1-e^{2}\right) ^{7/2}}  \label{b25}
\end{equation}
and 
\begin{equation}
g\left( e\right) =\frac{\left( 1+\frac{13}{4}e^{2}+\frac{7}{16}e^{4}\right) 
}{\left( 1-e^{2}\right) ^{7/2}}\text{ \smallskip ,}  \label{b26}
\end{equation}
while $M$ is the total mass of the binary and $\mu $ its reduced mass.

These parameters have to be determined in terms of the predictions of the
tensor-scalar theory. The correction, relatively to the predictions of General Relativity,
thereby introduced in the tensor part of the radiation will be of the order of the scalar
coupling $a_{BD}^{2}$, the value of which has to be inferior to $10^{-3}$ according to solar system
experiments. However, the ratio of the predicted rates of decrease in the orbital period
only depends on the eccentricity of the orbit (assumed to be keplerian), the value
of which is determined independently of the underlying theory. Hence, we may
readily understand what is the impact of the scalar component thanks to this ratio
(which is equal to the energy emission rates ratio, in the keplerian limit): 
\begin{equation}
\frac{d}{dt}T_{\left( h=0\right) }/\frac{d}{dt}T_{\left( h=2\right) }=\frac{d%
}{dt}E_{\left( h=0\right) }/\frac{d}{dt}E_{\left( h=2\right) }=\left(
a_{BD}^{2}/6\right) f\left( e\right) /g\left( e\right) \text{ .}  \label{b27}
\end{equation}
For an eccentricity of the order of $0.6$ and a scalar coupling fixed at $%
a_{BD}^{2}\simeq 10^{-3}$, we find 
\begin{equation}
\frac{d}{dt}T_{\left( h=0\right) }/\frac{d}{dt}T_{\left( h=2\right) }\mid
_{1913+16}\preccurlyeq 2\text{ }10^{-4}\text{ .}  \label{b28}
\end{equation}
The observed rate of decrease in the orbital period of the binary (\cite{damour},\cite{damour2}),
consistent with the predictions of General Relativity is given by
\begin{equation}
\frac{d}{dt}T_{obs}\mid _{1913+16}=-2.422(6)\text{ }10^{-12}s/s\text{ ,}
\label{b29}
\end{equation}
where the figure in parentheses represents a $1{\sigma}$ uncertainty in the last quoted digit.
All corrections applied, the precision of this measurement reaches a value of
$3.5\text{ }10^{-3}$.
Hence, the weakness of the scalar corrections just introduced suggests that it is not possible to make any conclusion
relative to a potential scalar gravitational emission, and, thereby, discriminate between General Relativity and
tensor-scalar models through the analysis of the Hulse-Taylor binary system.

Note that this issue has already been studied in much more details in ref.
(\cite{damour2},\cite{damour3},\cite{damour4}) where it has been shown indeed,
that the predictions of Brans-Dicke theories fall, as well as those of General Relativity, within experimental error bars.
However, the point we wanted to emphasize here is that the formalism defined above may be applied to render more
transparent some results on this subject.

\section{Conclusion}

All theories of gravitation enjoy the gauge freedom associated to the
arbitrariness in the choice of coordinate system locally parametrizing the
spacetime manifold. The main point of this letter is to emphasize the
interest of \emph{different} choices for \emph{different} purposes,
particularly in the context of tensor-scalar theories. On the one hand, a
comobile frame is always (whatever the theory) most appropriate when
exploring the impact of a Gravitational Wave impinging on a system of masses
and, specifically, to analyze the reaction of a Gravitational Wave detector.
On the other hand, the choice of the so-called $\Theta $TT-gauge allows the
decoupling of the tensor and scalar components of radiation, which is of
great interest when comparing predictions from tensor-scalar theories and
General Relativity. This is shown explicitly in the last two sections, in which our
formalism is applied to the assessment of the scalar contribution to the decrease
rate in the orbital period for a binary system.

\section{Acknowledgements}

We are grateful to M. Buysse, J.M. G\'{e}rard, J. Govaerts, C.
Smith and J. Weyers for interesting discussions and encouragements. This
work, part of a Master's thesis presented in June 1999, is supported by the
Fonds National de la Recherche Scientifique (FNRS), Belgium.\pagebreak 

\QTP{bibsection}


\begin{thebibliography}{9}
\bibitem{green}  M. Green, J. Schwarz and Ed. Witten, \emph{Superstring Theory%
}, \textbf{Vol. I}, (Cambridge University Press, Cambridge, 1987).

\bibitem{weinberg}  S. Weinberg, \emph{Gravitation and Cosmology}, (John
Wiley \& Sons, New York, 1972).

\bibitem{schutz}  B.F. Schutz, \emph{A first course in general relativity},
(Cambridge University Press, Cambridge, 1990).

\bibitem{fucito}  M. Bianchi, M. Brunetti, E. Coccia, F. Fucito and J.A.
Lobo, \emph{Cross section of a Resonant-Mass Detector for Scalar
Gravitational Waves}, Phys. Rev. \textbf{D57} (1998) 4525-4534.

\bibitem{maggiore}  M. Maggiore and A. Nicolis, \emph{Detection strategies
for scalar gravitational waves with interferometers and resonant spheres},
preprint, gr-qc/9907055 (1999).

\bibitem{mohanty}  S.\ Mohanty et P.K.\ Panda, \emph{Particle Physics Bounds
from the Hulse-Taylor Binary}, Phys. Rev. \textbf{D53} (1996) 5723-5726.

\bibitem{peters}  P.C. Peters and J. Mathews, \emph{Gravitational Radiation
from Point Masses in a Keplerian Orbit}, Phys. Rev. \textbf{Vol. 131, 1 }%
(1963) 435-440.

\bibitem{damour}  T. Damour, \emph{Experimental Tests of Relativistic Gravity%
}, preprint , gr-qc/9904057 (1999).

\bibitem{damour2}  T. Damour and G. Esposito-Far{\`e}se, \emph{Tensor-scalar gravity and
binary pulsar experiments}, Phys. Rev.
 \textbf{D54} (1996) 1474-1491.


\bibitem{damour3}  T. Damour and G. Esposito-Far{\`e}se, \emph{Tensor multiscalar theories of gravitation}, Class. Quant. Grav.
 \textbf{9} (1992) 2093-2176.

\bibitem{damour4}  T. Damour and J.H. Taylor, \emph{On the orbital period change of the binary pulsar PSR1913+16},
Astrophys. J. \textbf{366} (1991) 501-511.

 
\end{thebibliography}
\end{document}